\documentclass[conference]{IEEEtran}
\IEEEoverridecommandlockouts

\usepackage{bm}
\usepackage{cite}
\usepackage{xcolor}
\usepackage{amsthm}
\usepackage{amsmath}
\usepackage{amssymb}
\usepackage{amsfonts}
\usepackage{arydshln}
\usepackage{graphicx}
\usepackage{setspace}
\usepackage{algorithm}
\usepackage{algorithmic}

\newcounter{MYtempeqncnt}

\newtheorem*{lemma*}{Lemma}

\newtheorem*{theorem*}{Theorem}

\def\BibTeX{{\rm B\kern-.05em{\sc i\kern-.025em b}\kern-.08em
    T\kern-.1667em\lower.7ex\hbox{E}\kern-.125emX}}

\begin{document}

\title{Reconfigurable Intelligent Surface Aided \\ Wireless Localization
}

\author{\IEEEauthorblockN{Yiming Liu, Erwu Liu and Rui Wang}
	\IEEEauthorblockA{\textit{College of Electronics and Information Engineering, Tongji University, Shanghai, China} \\
		Emails: ymliu\_970131@tongji.edu.cn, erwu.liu@ieee.org, ruiwang@tongji.edu.cn}
}

\maketitle

\begin{abstract}
The advantages of millimeter-wave and large antenna arrays technologies for accurate wireless localization received extensive attentions recently. However, how to further improve the accuracy of wireless localization, even in the case of obstructed line-of-sight, is largely undiscovered. In this paper, the reconfigurable intelligent surface (RIS) is introduced into the system to make the positioning more accurate. First, we establish the three-dimensional RIS-assisted wireless localization channel model. After that, we derive the Fisher information matrix and the Cramer-Rao lower bound for evaluating the estimation of absolute mobile station position. Finally, we propose an alternative optimization method and a gradient decent method to optimize the reflect beamforming, which aims to minimize the Cramer-Rao lower bound to obtain a more accurate estimation. Our results show that the proposed methods significantly improve the accuracy of positioning, and decimeter-level or even centimeter-level positioning can be achieved by utilizing the RIS with a large number of reflecting elements.
\end{abstract}


\section{Introduction}
The fifth generation (5G) wireless technology offers great opportunities for accurate wireless localization due to the very high carrier frequency and large antenna arrays. Moreover, the sixth generation (6G) systems continue to develop towards even higher frequency ranges, \textit{e.g.}, at the millimeter-wave (mm-wave) as well as Tera Hertz (THz) ranges. The 6G white paper on localization and sensing released by University of Oulu \cite{2006.01779} points out that 6G systems will not only provide ubiquitous communication but also empower high accuracy localization due to the very fine angular resolutions. Decimeter-level or even centimeter-level positioning will be achieved by taking advantages of the new 6G key technology enablers. 

The advantages of mm-wave and large antenna arrays for accurate wireless localization have been extensively studied. Jeong, \textit{et al.}, propose an active beamforming method to enhance the localization accuracy of distributed antenna systems \cite{6799306}. Shahmansoori, \textit{et al.}, study the position and orientation estimation through mm-wave massive multiple-input multiple-output (MIMO) systems \cite{8240645}. Wang, Wu, and Shen prove the asymptotic spatial orthogonality of large-scale MIMO localization with general non-orthogonal waveforms \cite{8523631}. They also derive the Fisher information matrices for the position and orientation to characterize the performance bounds of MIMO localization. Zhou, \textit{et al.}, focus on active beamforming to reduce the localization error, and propose a successive localization and beamforming scheme \cite{8624270}.

However, how to further improve the accuracy of wireless localization, even in the case of obstructed line-of-sight, is largely undiscovered. Thus, in this paper, we introduce the reconfigurable intelligent surface (RIS), which is an efficient method to control the wavefront of the impinging signals, into the system to make the positioning more accurate and energy efficient. Our prior work shows that an RIS-assisted wireless communication system can achieve both high spectral and energy efficiency \cite{Liu2012:Energy}. Wu and Zhang verify that the RIS is able to drastically enhance the link quality over the conventional setup without the RIS \cite{8811733}.

In this paper, we introduce the RIS into the wireless localization system, and focus on the (passive) reflect beamforming to reduce the localization error. We first build the three-dimensional RIS-assisted wireless localization system model. After that, we derive the Fisher information matrix and the Cramer-Rao lower bound for the estimate of absolute mobile station (MS) position. For the sake of convenience in reflect beamforming design, we separate the response of RIS in the Fisher information matrix. Finally, we propose an alternative optimization method and a gradient decent method to optimize the reflect beamforming, which aims to minimize the Cramer-Rao lower bound. Our results show that the proposed methods significantly improve the accuracy of positioning, and decimeter-level or even centimeter-level positioning can be achieved by utilizing the RIS with a large number of reflecting elements.

\section{System and Channel Model}
Encouraged by the potential of mm-wave signals and large antenna arrays to improve the accuracy of positioning, many standardized channel models have started to emerge \cite{8240645,7173440,8624270}. We base the evaluation of our work on these models while introducing the RIS into the system to make the MS positioning more accurate. Introducing RIS can provide extra reflection paths which help in the estimation of MS position. On the one hand, it makes the positioning become possible when the line-of-sight is obstructed; On the other hand, it adds more position information for estimation.

\subsection{System Model}
We consider an RIS-assisted wireless localization system as illustrated in Fig. 1, where the base station (BS) is equipped with $N_{\mathrm{t}}$ antennas, the MS is equipped with $N_{\mathrm{r}}$ antennas, and the RIS is equipped with $N$ reflecting elements. The BS and MS are placed in the horizontal plane, along the $y$-axis direction, and the RIS is placed in the vertical plane. Although the BS and MS are equipped with a large number of antennas, they can fit within the compact form because of the small wavelength of mm-wave. Thus, they can be viewed as two points, and the positions of them are respectively denoted as $\boldsymbol{p} = [p_x, p_y, 0]^{\mathrm{T}}$ and $\boldsymbol{q} = [q_x, q_y, 0]^{\mathrm{T}}$. Compared with the BS and MS, the RIS has a much larger size, thus each reflecting element needs to be considered separately. The position of the $i$th reflecting element is denoted as $\boldsymbol{s}^{i} = [s^{i}_x, s^{i}_y, s^{i}_z]^{\mathrm{T}}$. The values of $\boldsymbol{p}$ and $\boldsymbol{s}^{i}$ are assumed to be known, while the value of $\boldsymbol{q}$ is unknown and requires to be estimated.
\begin{figure}[htbp]
	\centerline{\includegraphics[width = 8.75 cm]{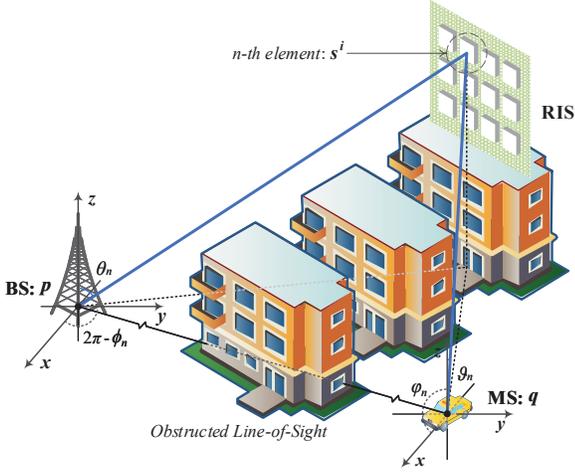}}
	\caption{The RIS-assisted wireless localization system with an $N_{\mathrm{t}}$-antenna BS, an $N_{\mathrm{r}}$-antenna MS, and an RIS comprising $N$ reflecting elements. This paper considers the case that the line-of-sight is obstructed by some blockages.}
	\label{fig 1}
\end{figure}

In this paper, we focus on the case that the line-of-sight is obstructed, thus there exist $N$ reflection paths through RIS. The elevation and azimuth angle-of-departure (AoD) of the $i$th path are denoted as $\theta_{i}$ and $\phi_{i}$. The elevation and azimuth angle-of-arrival (AoA) of the $i$th path are denoted as $\vartheta_{i}$ and $\varphi_{i}$. Due to that the positions of BS and RIS are fixed and known, we can obtain the values of $\theta_{i}$ and $\phi_{i}$ by means of the geometrical relationship between them. The values of $\vartheta_{i}$ and $\varphi_{i}$ are unknown and requires to be estimated. In addition, position estimate is equivalent with the azimuth and elevation AoA estimate due to the geometrical relationship.

\subsection{Channel Model}
We ignore the bounce reflections from the ground or other scatterers, since such paths get attenuated much more significantly than the paths through RIS. Based on the system model given above, the $N_{\mathrm{r}} \times N_{\mathrm{t}}$ channel matrix can be expressed as
\begin{equation}
	\tilde{\mathbf{H}} = \mathbf{\Lambda}_{\mathrm{r}} (\mathbf{H} \mathbf{\Phi}) \mathbf{\Lambda}_{\mathrm{t}}^{\mathrm{H}} \;\text{.}
\end{equation}
where the matrices $\mathbf{\Lambda}_{\mathrm{t}}$ and $ \mathbf{\Lambda}_{\mathrm{r}}$ are the array response matrices at BS and MS, the diagonal matrix $\mathbf{H}$ is the propagation gain matrix of $N$ paths, and the diagonal matrix $\mathbf{\Phi}$ represents the operations at the RIS.  The array response matrices $\mathbf{\Lambda}_{\mathrm{t}}$ and $ \mathbf{\Lambda}_{\mathrm{r}}$ depend on the angular parameters, given by
\begin{equation}
	\mathbf{\Lambda}_{\mathrm{t}} = \left[\mathbf{a}_{\mathrm{t}}\left(\theta_{1}, \phi_{1}\right), \cdots, \mathbf{a}_{\mathrm{t}}\left(\theta_{N}, \phi_{N}\right)\right] \in \mathbb{C}^{N_{\mathrm{t}} \times N} \;\;\text{,}
\end{equation}
\begin{equation}
\boldsymbol{\Lambda}_{\mathrm{r}}=\left[\mathbf{a}_{\mathrm{r}}\left(\vartheta_{1}, \varphi_{1}\right), \cdots, \mathbf{a}_{\mathrm{r}}\left(\vartheta_{N}, \varphi_{N}\right)\right] \in \mathbb{C}^{N_{\mathrm{r}} \times N} \;\text{,}
\end{equation}
where the $i^{\text{\textit{th}}}$ column vectors $\mathbf{a}_{\mathrm{t}}\left(\theta_{i}, \phi_{i}\right)$ and $\mathbf{a}_{\mathrm{r}}\left(\vartheta_{i}, \varphi_{i}\right)$ are
\begin{equation}
\mathbf{a}_{\mathrm{t}}\left(\theta_{i}, \phi_{i}\right) = [e^{j(1-1)k\sin\theta_{i}\sin\phi_{i}}, \cdots, e^{j(N_{\mathrm{t}}-1)k\sin\theta_{i}\sin\phi_{i}}]^{\mathrm{T}} \;\text{,}
\end{equation}
\begin{equation}
\mathbf{a}_{\mathrm{r}}\left(\vartheta_{i}, \varphi_{i}\right) = [e^{j(1-1)k\sin\vartheta_{i}\sin\varphi_{i}}, \cdots, e^{j(N_{\mathrm{r}}-1)k\sin\vartheta_{i}\sin\varphi_{i}}]^{\mathrm{T}}\text{.}
\end{equation}
The parameter $k = 2 \pi d / \lambda$ where $d$ is the separation between transmit/receive antennas at BS/MS, and $\lambda$ is the wavelength of transmitted signal. The diagonal matrix $\mathbf{\Phi}= \operatorname{diag}[e^{j\textcolor{black}{\boldsymbol{\varrho}}}]$ where $e^{j\textcolor{black}{\boldsymbol{\varrho}}}$ is an element-wise power operation, and the $N \times 1$ vector $\textcolor{black}{\boldsymbol{\varrho}=[\varrho_1, \cdots, \varrho_N]^{\mathrm{T}}}$ represents the phase shifts of $N$ reflecting elements at RIS. The diagonal matrix $\mathbf{H} = \operatorname{diag}[\mathbf{h}]$ where the $N \times 1$ vector $\mathbf{h} = [h_1, \cdots, h_N]^{\mathrm{T}}$ represents the propagation gains of $N$ reflection paths, and the entries in it are independent and identically distributed.

\subsection{Received Signal Model}
The signal transmitted by the BS is denoted as $\mathbf{X} \in \mathbb{C}^{N_{\mathrm{t}} \times L}$ where $L$ represents the number of consumed time slots, and the column vector $\boldsymbol{x}(i)$ in it represents the signal transmitted at the $i$th time slot. Applying the vectorization operation, the received signal $\boldsymbol{y} \in \mathbb{C}^{L N_{\mathrm{r}} \times 1}$ at the MS can be expressed as 
\begin{equation}
\boldsymbol{y} = \left[\begin{array}{c}
\boldsymbol{y}(1) \\
\boldsymbol{y}(2) \\
\vdots \\
\boldsymbol{y}(L)
\end{array}\right]=\left[\begin{array}{c}
\tilde{\mathbf{H}} \boldsymbol{x}(1) \\
\tilde{\mathbf{H}} \boldsymbol{x}(2) \\
\vdots \\
\tilde{\mathbf{H}} \boldsymbol{x}(L)
\end{array}\right]+\left[\begin{array}{c}
\boldsymbol{n} \\
\boldsymbol{n} \\
\vdots \\
\boldsymbol{n}
\end{array}\right] \;\text{,}
\end{equation}
where the vector $\boldsymbol{n} \in \mathbb{C}^{N_{\mathrm{r}} \times 1} $ is an additive white Gaussian noise with the elements independently drawn from $\mathcal{C}\mathcal{N} (0, \sigma^{2})$, and the transmit power at the $l$th slot is $p_{\mathrm{BS}} = \mathbb{E}\lbrace{\boldsymbol{x}^{\mathrm{H}}(l)}\boldsymbol{x}(l)\rbrace $.

\section{Cramer-Rao Bound on Position Estimation}
Based on the channel model, the unknown parameters to be estimated can be denoted as a $3N$ dimensional vector $\boldsymbol{\eta}$,
\begin{equation}
\boldsymbol{\eta} = [\vartheta_{1}, \cdots, \vartheta_{N}, \; \varphi_{1}, \cdots, \varphi_{N}, \; h_{1}, \cdots, h_{N}]^{\mathrm{T}} \; \text{.}
\end{equation}
We denote the unbiased estimator of $\boldsymbol{\eta}$ as $\hat{\boldsymbol{\eta}}$, and based on the Cramer-Rao theorem \cite{FSSPET,SSPDET}, the mean-square error is bounded as
\begin{equation}
	\mathbb{E}\left[(\hat{\boldsymbol{\eta}} - \boldsymbol{\eta})(\hat{\boldsymbol{\eta}} - \boldsymbol{\eta})^{\mathrm{H}}\right] \geq \mathbf{J}_{\boldsymbol{\eta}}^{-1} \;\text{,}
\end{equation}
where $\mathbf{J}_{\boldsymbol{\eta}}$ is the $3N \times 3N$ Fisher information matrix, and $[\mathbf{J}_{\boldsymbol{\eta}}^{-1}]_{m,m}$ is the Cramer-Rao lower bound for the $m$th parameter estimate. The $(m,n)$th entry of $\mathbf{J}_{\boldsymbol{\eta}}$ is defined as 
\begin{equation}
\label{Fisher information matrix 1}
\begin{aligned}
\left[\mathbf{J}_{\boldsymbol{\eta}}\right]_{m,n} & \triangleq \mathbb{E}\left[-\frac{\partial^{2} \ln p(\boldsymbol{y} ; \boldsymbol{\eta})}{\partial {\eta}_{m} \partial {\eta}_{n}}\right] \\
& \equiv \mathbb{E}\left[\frac{\partial \ln p(\boldsymbol{y} ; \boldsymbol{\eta})}{\partial {\eta}_{m}} \frac{\partial \ln p(\boldsymbol{y} ; \boldsymbol{\eta})}{\partial {\eta}_{n}}\right] \;\text{,}
\end{aligned}
\end{equation}
where $p(\boldsymbol{y} ; \boldsymbol{\eta})$ is the likelihood function of the random vector $\boldsymbol{y}$ conditioned on $\boldsymbol{\eta}$, and $\eta_{m}$ is the $m$th entry of $\boldsymbol{\eta}$. The proof of the identity in Eq. (\ref{Fisher information matrix 1}) is given in \cite{SSPDET}. Based on the received signal model, $p(\boldsymbol{y} ; \boldsymbol{\eta})$ can be written as 
\begin{equation}
	p(\boldsymbol{y} ; \boldsymbol{\eta}) = (2 \pi)^{-\frac{L N_{\mathrm{r}}}{2}}\left(\operatorname{det} \sigma^{2} \mathbf{I}\right)^{-\frac{1}{2}} e^{-\frac{1}{2}\left[\boldsymbol{\mu}^{\mathrm{H}}\left(\sigma^{2} \mathbf{I}\right)^{-1} \boldsymbol{\mu}\right]} \; \text{,}
\end{equation}
where the vector $\boldsymbol{\mu}$ is the mean vector of the random vector $\boldsymbol{y}$ conditioned on $\boldsymbol{\eta}$. 

To further obtain the Fisher information matrix $\mathbf{J}_{\boldsymbol{\eta}}$, we first present the following Lemma from Eq. (6) in \cite{1519691}:
\begin{lemma*}
	For an $N$ dimensional random vector $\boldsymbol{y}$ which obeys the symmetric complex Gaussian distribution $\mathcal{C}\mathcal{N} (\boldsymbol{\mu}, \mathbf{\Sigma})$, the $(m,n)$th entry of the Fisher information matrix is given by
	\begin{equation}
	\left[\mathbf{J}\right]_{m,n}=2 \mathfrak{Re} \left\{\frac{\partial \boldsymbol{\mu}^{\mathrm{H}}}{\partial \eta_{m}} \mathbf{\Sigma}^{-1} \frac{\partial \boldsymbol{\mu}}{\partial \eta_{n}}\right\} + \mathfrak{Tr}\left\lbrace \mathbf{\Sigma}^{-1} \frac{\partial \mathbf{\Sigma}}{\partial \eta_{m}} \mathbf{\Sigma}^{-1} \frac{\partial \mathbf{\Sigma}}{\partial \eta_{n}} \right\rbrace. 
	\end{equation}	
\end{lemma*}

By using the above Lemma, the $(m,n)$th entry of $\mathbf{J}_{\boldsymbol{\eta}}$ in Eq. (\ref{Fisher information matrix 1}) can be rewritten as
\begin{equation}
\label{Gaussian FIM}
\left[\mathbf{J}_{\boldsymbol{\eta}}\right]_{m,n} = \sum_{l=1}^{L} \frac{2}{\sigma^{2}} \mathfrak{Re} \left\{\frac{\partial {\boldsymbol{\mu}^{\mathrm{H}}(l)}}{\partial \eta_{m}} \frac{\partial \boldsymbol{\mu}(l)}{\partial \eta_{n}}\right\} = \sum_{l=1}^{L} \left[\mathbf{J}_{\boldsymbol{\eta}}(l)\right]_{m,n} \; \text{,}
\end{equation}	
where the mean vector $\boldsymbol{\mu}(l) = \tilde{\mathbf{H}} \boldsymbol{x}(l)$. Eq. (\ref{Gaussian FIM}) demonstrates that the Fisher information matrix $\mathbf{J}_{\boldsymbol{\eta}}$ is the sum of the Fisher information matrix $\mathbf{J}_{\boldsymbol{\eta}}(l)$ at each time slot. The rest in this section gives the solving process of $\mathbf{J}_{\boldsymbol{\eta}}(l)$.

We rewrite the Fisher information matrix $\mathbf{J}_{\boldsymbol{\eta}}(l)$ as a partitioned matrix in Eq. (\ref{partitioned matrix}) which is for later use.
\begin{equation}
\label{partitioned matrix}
\mathbf{J}_{\boldsymbol{\eta}}(l) = \frac{2}{\sigma^{2}}
\begin{array}{c@{}l}
	\left[
	\begin{array}{ccc;{2pt/2pt}c}
		\quad & \quad & \quad & \quad \\
		\quad & \bm{\mathcal{A}} & \quad &\bm{\mathcal{B}} \\
		\quad & \quad & \quad & \quad \\
		\hdashline[2pt/2pt]
		\quad & \quad & \quad & \quad \\
		\quad & \bm{\mathcal{C}} & \quad &\bm{\mathcal{D}} \\
	\end{array}
	\right]
	&\begin{array}{l}
		\left.\rule{-1.75 mm}{7.8 mm}\right\} {\scriptsize \text{2\textit{N}}}\\
		\left.\rule{-1.75 mm}{4.85 mm}\right\} {\scriptsize \text{\textit{\;N}}}
	\end{array}\\[-5pt]
	\begin{array}{cc}
		\underbrace{\rule{20.7 mm}{0 mm}}_{\text{2\textit{N}}}
		\underbrace{\rule{3.75 mm}{0 mm}}_{\text{\textit{N}}}
	\end{array}
\end{array}
\end{equation}
where the entries in submatrix $\bm{\mathcal{A}}$ contain the partial derivatives with respect to the azimuth/elevation AoA, while the entries in submatrices $\bm{\mathcal{B}}$, $\bm{\mathcal{C}}$ and $\bm{\mathcal{D}}$ contain the partial derivatives with respect to the propagation gains.

Next, we give the partial derivatives of $\boldsymbol{\mu}(l)$ with respect to the azimuth/elevation AoA. Before giving it, we first give the partial derivatives of the matrix $\mathbf{\Lambda}_{\mathrm{r}} (\mathbf{H} \mathbf{\Phi}) \mathbf{\Lambda}_{\mathrm{t}}^{\mathrm{H}}$. The $(m,n)$th entry of the matrix $\mathbf{\Lambda}_{\mathrm{r}} (\mathbf{H} \mathbf{\Phi}) \mathbf{\Lambda}_{\mathrm{t}}^{\mathrm{H}}$ is given by
\begin{equation}
\left[\mathbf{\Lambda}_{\mathrm{r}} (\mathbf{H} \mathbf{\Phi}) \mathbf{\Lambda}_{\mathrm{t}}^{\mathrm{H}}\right]_{m, n} = \sum_{i=1}^{N} h_{i} e^{j\left[\textcolor{black}{\varrho_{i}}+(m-1)\xi_{i}+(1-n)\zeta_{i}\right]} \;\text{,}
\end{equation}
where $\xi_{i} = k \sin \vartheta_{i} \sin \varphi_{i} $ and $\zeta_{i} = k \sin \theta_{i} \sin \phi_{i}$. To separate the response of RIS, the partial derivatives of the matrix $\mathbf{\Lambda}_{\mathrm{r}} (\mathbf{H} \mathbf{\Phi}) \mathbf{\Lambda}_{\mathrm{t}}^{\mathrm{H}}$ with respect to the azimuth/elevation AoA are given by
\begin{equation}
\frac{\partial\left[\mathbf{\Lambda}_{\mathrm{r}}(\mathbf{H} \mathbf{\Phi}) \mathbf{\Lambda}_{\mathrm{t}}^{\mathrm{H}}\right]}{\partial \vartheta_{i}} = e^{j \textcolor{black}{\varrho_{i}}} \boldsymbol{\dot{\Xi}}_{i} \in \mathbb{C}^{N_{\mathrm{r}} \times N_{\mathrm{t}}} \;\text{,}
\end{equation}
\begin{equation}
\frac{\partial\left[\mathbf{\Lambda}_{\mathrm{r}}(\mathbf{H} \mathbf{\Phi}) \mathbf{\Lambda}_{\mathrm{t}}^{\mathrm{H}}\right]}{\partial \varphi_{i}} = e^{j \textcolor{black}{\varrho_{i}}} \boldsymbol{\ddot{\Xi}}_{i} \in \mathbb{C}^{N_{\mathrm{r}} \times N_{\mathrm{t}}} \;\text{,}
\end{equation}
where
\begin{equation}
[\boldsymbol{\dot{\Xi}}_{i}]_{m,n} =h_{i} j (m-1) k \sin \varphi_{i} \cos \vartheta_{i} e^{j\left[(m-1)\xi_{i}+(1-n)\zeta_{i}\right]} \;\text{,}
\end{equation}
\begin{equation}
[\boldsymbol{\ddot{\Xi}}_{i}]_{m,n} =h_{i} j (m-1) k \sin \vartheta_{i} \cos \varphi_{i} e^{j\left[(m-1)\xi_{i}+(1-n)\zeta_{i}\right] } \;\text{.}
\end{equation}

Then, we obtain the partial derivatives of $\boldsymbol{\mu}(l)$ with respect to the azimuth/elevation AoA as follows:
\begin{equation}
\label{partial derivatives with respect to the azimuth AoA}
\frac{\partial \boldsymbol{\mu}(l)}{\partial \vartheta_{i}} = e^{j \textcolor{black}{\varrho_{i}}} \boldsymbol{\dot{\varpi}}_{i} (l) \in \mathbb{C}^{N_{\mathrm{r}} \times 1} \;\text{,}
\end{equation}
\begin{equation}
\label{partial derivatives with respect to the elevation AoA}
\frac{\partial \boldsymbol{\mu}(l)}{\partial \varphi_{i}} = e^{j \textcolor{black}{\varrho_{i}}} \boldsymbol{\ddot{\varpi}}_{i} (l) \in \mathbb{C}^{N_{\mathrm{r}} \times 1} \;\text{,}
\end{equation}
where $\boldsymbol{\dot{\varpi}}_{i}(l) = \boldsymbol{\dot{\Xi}}_{i} \boldsymbol{x}(l)$ and $\boldsymbol{\ddot{\varpi}}_{i}(l) = \boldsymbol{\ddot{\Xi}}_{i} \boldsymbol{x}(l)$.

Compared with the azimuth/elevation AoA, what we are more interested in is the absolute MS position $\boldsymbol{q}_{1:2} = [q_{x}, q_{y}]^{\mathrm{T}}$. The Fisher information matrix of it can be obtained by means of the $2 \times 3N$ transformation matrix $\mathbf{T}$, which is expressed as 
\begin{equation}
\label{FIM for position}
\mathbf{J}_{\boldsymbol{q}}(l) = \mathbf{T} \mathbf{J}_{\boldsymbol{\eta}}(l) \mathbf{T}^{\mathrm{T}} \;\text{,}
\end{equation}
where the transformation matrix $\mathbf{T}$ is defined as
\begin{equation}
\mathbf{T} \triangleq \frac{\partial \boldsymbol{\eta}^{\mathrm{T}}}{\partial \boldsymbol{q}_{1:2}} \; \text{.}
\end{equation}
The entries of the transformation matrix $\mathbf{T}$ can be obtained by the geometrical relationship between the azimuth/elevation AoA and the MS position which is shown as follows:
\begin{equation}
\label{geometric relationship between the elevation AoA and the MS position}
\vartheta_{i} = \arctan \left[\frac{\left\|\boldsymbol{q}_{1:2}-\boldsymbol{s}_{1: 2}^{i}\right\|_{2}}{s_{z}^{i}}\right] \; \text{,}
\end{equation}
\begin{equation}
\label{geometric relationship between the azimuth AoA and the MS position}
\varphi_{i}=\arccos \left[-\frac{\left|q_{x}-s_{x}^{i}\right|\;}{\left\|\boldsymbol{q}_{1:2}-\boldsymbol{s}_{1: 2}^{i}\right\|_{2}}\right] \; \text{.}
\end{equation}

Then, by computing the partial derivatives of the azimuth and elevation AoA with respect to the MS position, we obtain the transformation matrix $\mathbf{T}$ as follows:
\begin{equation}
\begin{array}{cc}
\mathbf{T}=
\left[
\begin{array}{ccc;{2pt/2pt}ccc;{2pt/2pt}ccc}
\;&\bm{\mathcal{E}}&\;&  \;&\bm{\mathcal{F}}&\;&  \;&\bm{\mathcal{G}}\;&
\end{array}
\right]

\\[-6pt]
\begin{array}{cccc}
\;\;\;\;\;\;\;
\underbrace{\rule{15.25 mm}{0 mm}}_{\text{\textit{N}}}
\underbrace{\rule{15.2 mm}{0 mm}}_{\text{\textit{N}}}
\underbrace{\rule{15.2 mm}{0 mm}}_{\text{\textit{N}}}
\end{array}
\end{array}
\end{equation}
where the $i$th column vector of the submatrices $\bm{\mathcal{E}}$, $\bm{\mathcal{F}}$ and $\bm{\mathcal{G}}$ respectively are
\begin{equation}
\label{submatrix_E}
[\bm{\mathcal{E}}]_{i} = \frac{\partial \vartheta_{i}}{\partial \boldsymbol{q}_{1:2}} = \frac{s_{z}^{i}}{\left\|\boldsymbol{q}_{1:2}-\boldsymbol{s}_{1: 2}^{i}\right\|_{2}^{2} + {(s_{z}^{i})}^{2}} [-\cos \varphi_{i}, \sin \varphi_{i}]^{\mathrm{T}} \; \text{,}
\end{equation}
\begin{equation}
\label{submatrix_F}
[\bm{\mathcal{F}}]_{i} =  \frac{\partial \varphi_{i}}{\partial \boldsymbol{q}_{1:2}} = \frac{1}{\left\|\boldsymbol{q}_{1:2}-\boldsymbol{s}_{1: 2}^{i}\right\|_{2}} [\sin \varphi_{i}, \cos \varphi_{i}]^{\mathrm{T}} \; \text{,} \qquad \qquad\; 
\end{equation}
\begin{equation}
\label{submatrix_G}
[\bm{\mathcal{G}}]_{i} = \frac{\partial h_{i}}{\partial \boldsymbol{q}_{1:2}} = [0, 0]^{\mathrm{T}} \;\text{.}  \; \qquad \qquad \qquad \qquad \qquad \qquad \qquad 
\end{equation}
The Eqs. (\ref{submatrix_E})\;-\;(\ref{submatrix_G}) are derived in subsection A of Appendix. Because the $2 \times N$ submatrix $\bm{\mathcal{G}}$ is a zero matrix, then based on the multiplication principle of partitioned matrix, Eq. (\ref{FIM for position}) can be simplified as 
\begin{equation}
\mathbf{J}_{\boldsymbol{q}}(l) = \frac{2}{\sigma^{2}} \tilde{\mathbf{T}} \bm{\mathcal{A}} \tilde{\mathbf{T}}^{\mathrm{T}} \;\text{,}
\end{equation}
where the matrix $\tilde{\mathbf{T}}$ consists of the submatrices $\bm{\mathcal{E}}$ and $\bm{\mathcal{F}}$, and the matrix $\bm{\mathcal{A}}$ is the $2N \times 2N$ submatrix in Eq. (\ref{partitioned matrix}). 

We rewrite the matrix $\bm{\mathcal{A}}$ as a partitioned matrix as follows:
\begin{equation}
\begin{array}{c@{}l}
\bm{\mathcal{A}} = 
\left[
\begin{array}{c;{2pt/2pt}c}
\quad & \quad\\
\quad \bm{\mathcal{H}} \quad & \quad \bm{\mathcal{I}} \quad \\
\quad & \quad \\
\hdashline[2pt/2pt]
\quad & \quad\\
\quad \bm{\mathcal{J}} \quad & \quad\bm{\mathcal{K}} \quad \\
\quad & \quad \\
\end{array}
\right]
\begin{array}{l}
\left.\rule{-2.5 mm}{6.75 mm}\right\} {\scriptsize \text{\textit{N}}}\\
\left.\rule{-2.5 mm}{6.75 mm}\right\} {\scriptsize \text{\textit{N}}}
\end{array}\\[-5pt]
\begin{array}{ccc}
\;
\underbrace{\rule{14.5 mm}{0 mm}}_{\text{\textit{N}}}
\underbrace{\rule{14 mm}{0 mm}}_{\text{\textit{N}}}
\;
\end{array}
\end{array}
\end{equation}
Based on the definition of the $(m,n)$th entry of $\mathbf{J}_{\boldsymbol{\eta}}(l)$ in Eq. (\ref{Gaussian FIM}) and the partial derivatives of $\boldsymbol{\mu}(l)$ with respect to the azimuth/elevation AoA in Eqs. (\ref{partial derivatives with respect to the azimuth AoA}) and (\ref{partial derivatives with respect to the elevation AoA}), we find that the response of RIS in the $(m,n)$th entry of the submatrix $\bm{\mathcal{H}}$ is same with those in the submatrices $\bm{\mathcal{I}}$, $\bm{\mathcal{J}}$ and $\bm{\mathcal{K}}$, which is $e^{j(\textcolor{black}{\varrho_{n}}-\textcolor{black}{\varrho_{m}})}$. Then, the entries in the $2 \times 2$ Fisher information matrix $\mathbf{J}_{\boldsymbol{q}}(l)$ can be given as 
\begin{equation}
\label{J11}
\left[\mathbf{J}_{\boldsymbol{q}}(l)\right]_{1,1} = \frac{2}{\sigma^{2}} \sum_{m=1}^{N} \sum_{n=1}^{N} \mathfrak{Re} \left[e^{j(\textcolor{black}{\varrho_{n}}-\textcolor{black}{\varrho_{m}})} \kappa_{m, n}^{1,1}(l)\right] \;\text{,}
\end{equation}
\begin{equation}
\label{J12}
\left[\mathbf{J}_{\boldsymbol{q}}(l)\right]_{1,2} = \frac{2}{\sigma^{2}} \sum_{m=1}^{N} \sum_{n=1}^{N} \mathfrak{Re} \left[e^{j(\textcolor{black}{\varrho_{n}}-\textcolor{black}{\varrho_{m}})} \kappa_{m, n}^{1,2}(l)\right] \;\text{,}
\end{equation}
\begin{equation}
\label{J21}
\left[\mathbf{J}_{\boldsymbol{q}}(l)\right]_{2,1} = \frac{2}{\sigma^{2}} \sum_{m=1}^{N} \sum_{n=1}^{N} \mathfrak{Re} \left[e^{j(\textcolor{black}{\varrho_{n}}-\textcolor{black}{\varrho_{m}})} \kappa_{m, n}^{2,1}(l)\right] \;\text{,}
\end{equation}
\begin{equation}
\label{J22}
\left[\mathbf{J}_{\boldsymbol{q}}(l)\right]_{2,2} = \frac{2}{\sigma^{2}} \sum_{m=1}^{N} \sum_{n=1}^{N} \mathfrak{Re} \left[e^{j(\textcolor{black}{\varrho_{n}}-\textcolor{black}{\varrho_{m}})} \kappa_{m, n}^{2,2}(l)\right] \;\text{,}
\end{equation}
where the scalars $\kappa_{m, n}^{1,1}(l)$, $\kappa_{m, n}^{1,2}(l)$, $\kappa_{m, n}^{2,1}(l)$ and $\kappa_{m, n}^{2,2}(l)$ are
\begin{equation}
\begin{aligned}
\kappa_{m, n}^{1,1}(l) = & \; \alpha_{m}^{x} \alpha_{n}^{x} \boldsymbol{\dot{\varpi}}_{m}^{\mathrm{H}}(l) \boldsymbol{\dot{\varpi}}_{n}(l) + \beta_{m}^{x} \alpha_{n}^{x} \boldsymbol{\ddot{\varpi}}_{m}^{\mathrm{H}}(l) \boldsymbol{\dot{\varpi}}_{n}(l) \\ + & \; \alpha_{m}^{x} \beta_{n}^{x} \boldsymbol{\dot{\varpi}}_{m}^{\mathrm{H}}(l) \boldsymbol{\ddot{\varpi}}_{n}(l) + \beta_{m}^{x} \beta_{n}^{x} \boldsymbol{\ddot{\varpi}}_{m}^{\mathrm{H}}(l) \boldsymbol{\ddot{\varpi}}_{n}(l) \;\text{,}
\end{aligned}
\end{equation}
\begin{equation}
\begin{aligned}
\kappa_{m, n}^{1,2}(l) = & \; \alpha_{m}^{x} \alpha_{n}^{y} \boldsymbol{\dot{\varpi}}_{m}^{\mathrm{H}}(l)\boldsymbol{\dot{\varpi}}_{n}(l) + \beta_{m}^{x} \alpha_{n}^{y} \boldsymbol{\ddot{\varpi}}_{m}^{\mathrm{H}}(l) \boldsymbol{\dot{\varpi}}_{n}(l)\\ + & \; \alpha_{m}^{x} \beta_{n}^{y} \boldsymbol{\dot{\varpi}}_{m}^{\mathrm{H}}(l) \boldsymbol{\ddot{\varpi}}_{n}(l) + \beta_{m}^{x} \beta_{n}^{y} \boldsymbol{\ddot{\varpi}}_{m}^{\mathrm{H}}(l) \boldsymbol{\ddot{\varpi}}_{n}(l) \;\text{,}
\end{aligned}
\end{equation}
\begin{equation}
\begin{aligned}
\kappa_{m, n}^{2,1}(l) = & \; \alpha_{m}^{y} \alpha_{n}^{x} \boldsymbol{\dot{\varpi}}_{m}^{\mathrm{H}}(l) \boldsymbol{\dot{\varpi}}_{n}(l) + \beta_{m}^{y} \alpha_{n}^{x} \boldsymbol{\ddot{\varpi}}_{m}^{\mathrm{H}}(l) \boldsymbol{\dot{\varpi}}_{n}(l) \\ + & \; \alpha_{m}^{y} \beta_{n}^{x} \boldsymbol{\dot{\varpi}}_{m}^{\mathrm{H}}(l) \boldsymbol{\ddot{\varpi}}_{n}(l) + \beta_{m}^{y} \beta_{n}^{x} \boldsymbol{\ddot{\varpi}}_{m}^{\mathrm{H}}(l) \boldsymbol{\ddot{\varpi}}_{n}(l) \;\text{,}
\end{aligned}
\end{equation}
\begin{equation}
\begin{aligned}
\kappa_{m, n}^{2,2}(l) = & \; \alpha_{m}^{y} \alpha_{n}^{y} \boldsymbol{\dot{\varpi}}_{m}^{\mathrm{H}}(l) \boldsymbol{\dot{\varpi}}_{n}(l) + \beta_{m}^{y} \alpha_{n}^{y} \boldsymbol{\ddot{\varpi}}_{m}^{\mathrm{H}}(l) \boldsymbol{\dot{\varpi}}_{n}(l) \\ + & \; \alpha_{m}^{y} \beta_{n}^{y} \boldsymbol{\dot{\varpi}}_{m}^{\mathrm{H}}(l) \boldsymbol{\ddot{\varpi}}_{n}(l) + \beta_{m}^{y} \beta_{n}^{y} \boldsymbol{\ddot{\varpi}}_{m}^{\mathrm{H}}(l) \boldsymbol{\ddot{\varpi}}_{n}(l) \;\text{,}
\end{aligned}
\end{equation}
and the parameters $\alpha_{m}^{x}$, $\alpha_{m}^{y}$, $\beta_{m}^{x}$ and $\beta_{m}^{y}$ respectively are the first and second entries of $[\bm{\mathcal{E}}]_{m}$ and $[\bm{\mathcal{F}}]_{m}$. Then, by using Eq. (\ref{Gaussian FIM}), we obtain the entries of the Fisher information matrix $\mathbf{J}_{\boldsymbol{q}}$ as follows:
\begin{equation}
\label{eq. 39}
\left[\mathbf{J}_{\boldsymbol{q}}\right]_{a,b} = \sum_{l=1}^{L} \left[\mathbf{J}_{\boldsymbol{q}}(l)\right]_{a,b} \text{,} \; a = 1\text{,}\;2 ; \; b = 1\text{,}\;2 \; \text{.}
\end{equation}

Finally, we obtain the Cramer-Rao lower bound for the MS position estimate which is the trace of the inverse matrix of the Fisher information matrix $\mathbf{J}_{\boldsymbol{q}}$. The above result can be summarized as a theorem:

\begin{theorem*}
	For an RIS-assisted wireless localization system modeled as \textit{Section II}, the Cramer-Rao lower bound for the estimate of absolute MS position can be expressed as 
	\begin{equation}
	\text{CRLB} = \mathfrak{Tr}\left(\mathbf{J}_{\boldsymbol{q}}^{-1}\right) = \frac{\left[\mathbf{J}_{\boldsymbol{q}}\right]_{1,1} + \left[\mathbf{J}_{\boldsymbol{q}}\right]_{2,2}} {\left[\mathbf{J}_{\boldsymbol{q}}\right]_{1,1}\left[\mathbf{J}_{\boldsymbol{q}}\right]_{2,2} - \left[\mathbf{J}_{\boldsymbol{q}}\right]_{1,2}\left[\mathbf{J}_{\boldsymbol{q}}\right]_{2,1}} \; \text{.}
	\end{equation}
\end{theorem*}

\vspace{3pt}

\section{Reflect Beamforming Design at RIS}
In this paper, we aim to minimize the Cramer-Rao lower bound for the MS position estimate by optimizing the reflect beamforming at the RIS. Accordingly, the optimization problem can be formulated as 
\begin{equation}
\label{p1}
\begin{aligned}
\min_{\boldsymbol{\varrho}} &\quad f({\boldsymbol{\varrho}}) = \frac{\left[\mathbf{J}_{\boldsymbol{q}}\right]_{1,1} + \left[\mathbf{J}_{\boldsymbol{q}}\right]_{2,2}} {\left[\mathbf{J}_{\boldsymbol{q}}\right]_{1,1}\left[\mathbf{J}_{\boldsymbol{q}}\right]_{2,2} - \left[\mathbf{J}_{\boldsymbol{q}}\right]_{1,2}\left[\mathbf{J}_{\boldsymbol{q}}\right]_{2,1}} \\
\operatorname{s.t.} &\quad {0 \leq {\varrho_{n}} \leq 2 \pi, \quad \forall n=1, \cdots, N}.
\end{aligned}
\end{equation}
Because the objective function in Eq. (\ref{p1}) is a non-convex fractional function, it is difficult to obtain the global optimal solution. To further analyze this problem, we utilize the \textit{gradient decent method} (GDM) to make the reflect beamforming design. In the \textit{gradient flow direction} which is defined as the direction of the negative gradient $-\nabla_{\boldsymbol{\varrho}} f(\boldsymbol{\varrho})$, the function $f(\boldsymbol{\varrho})$ decreases at the maximum descent rate. And each component of the gradient vector gives the rate of change of the scalar function in the component direction  \cite{zhang2017matrix}. 

The function $f(\boldsymbol{\varrho})$
is a scalar function with respect to an $N \times 1$ vector-variable. Thus, the gradient of it is expressed as
\begin{equation}
\nabla_{{\boldsymbol{\varrho}}} f({\boldsymbol{\varrho}}) = \left[\frac{\partial f({\boldsymbol{\varrho}})}{\partial {\varrho_{1}}}, \frac{\partial f({\boldsymbol{\varrho}})}{\partial {\varrho_{2}}}, \cdots, \frac{\partial f({\boldsymbol{\varrho}})}{\partial {\varrho_{N}}}\right]^{\mathrm{T}} \;\text{,}
\end{equation}
where the partial derivative of $f(\boldsymbol{\varrho})$ with respect to $\varrho_{i}$ is given in Eq. (\ref{partial derivative with respect to each component}) at the top of the next page.
\begin{figure*}[!t]
	\normalsize
	\setcounter{MYtempeqncnt}{\value{equation}}
	\begin{equation}
	\label{partial derivative with respect to each component}
	\frac{\partial f(\boldsymbol{\varrho})}{\partial \varrho_{i}} = \frac{\left(\frac{\partial\left[\mathbf{J}_{\boldsymbol{q}}\right]_{1,1}}{\partial \varrho_{i}}+\frac{\partial\left[\mathbf{J}_{\boldsymbol{q}}\right]_{2,2}}{\partial \varrho_{i}}\right) \mathfrak{De}-\mathfrak{N}_{\mathfrak{u}}\left(\frac{\partial\left[\mathbf{J}_{\boldsymbol{q}}\right]_{1,1}}{\partial \varrho_{i}}\left[\mathbf{J}_{\boldsymbol{q}}\right]_{2,2}+\left[\mathbf{J}_{\boldsymbol{q}}\right]_{1,1} \frac{\partial\left[\mathbf{J}_{\boldsymbol{q}}\right]_{2,2}}{\partial \varrho_{i}}-\frac{\partial\left[\mathbf{J}_{\boldsymbol{q}}\right]_{1,2}}{\partial \varrho_{i}}\left[\mathbf{J}_{\boldsymbol{q}}\right]_{2,1}-\left[\mathbf{J}_{\boldsymbol{q}}\right]_{1,2} \frac{\partial\left[\mathbf{J}_{\boldsymbol{q}}\right]_{2,1}}{\partial \varrho_{i}}\right)}{{\mathfrak{De}}^2}
	\end{equation}
	\setcounter{figure}{\value{MYtempeqncnt}}
	\hrulefill
	\vspace*{4 pt}
\end{figure*} 
The parameters $\mathfrak{Nu}$ and $\mathfrak{De}$ in it represent the numerator and denominator of $f(\boldsymbol{\varrho})$, respectively, and the partial derivatives of the entries in the Fisher information matrix $\mathbf{J}_{\boldsymbol{q}}$ with respect to $\varrho_{i}$ is given as follows:
\begin{equation}
\label{partial derivative of J_1,1}
\frac{\partial \left[\mathbf{J}_{q}\right]_{1,1}}{\partial {\varrho_{i}}} = \frac{4}{\sigma^{2}} \sum_{l=1}^{L} \sum_{n \neq i}^{N} \mathfrak{Re} \left[je^{j(\textcolor{black}{\varrho_{i}}-\textcolor{black}{\varrho_{n}})} \kappa_{n, i}^{1,1}(l)\right] \;\text{,}
\end{equation}
\begin{equation}
\label{partial derivative of J_1,2}
\frac{\partial \left[\mathbf{J}_{q}\right]_{1,2}}{\partial {\varrho_{i}}} = \frac{4}{\sigma^{2}} \sum_{l=1}^{L} \sum_{n \neq i}^{N} \mathfrak{Re} \left[je^{j(\textcolor{black}{\varrho_{i}}-\textcolor{black}{\varrho_{n}})} \kappa_{n, i}^{1,2}(l)\right] \;\text{,}
\end{equation}
\begin{equation}
\label{partial derivative of J_2,1}
\frac{\partial \left[\mathbf{J}_{q}\right]_{2,1}}{\partial {\varrho_{i}}} = \frac{4}{\sigma^{2}} \sum_{l=1}^{L} \sum_{n \neq i}^{N} \mathfrak{Re} \left[je^{j(\textcolor{black}{\varrho_{i}}-\textcolor{black}{\varrho_{n}})} \kappa_{n, i}^{2,1}(l)\right] \;\text{,}
\end{equation}
\begin{equation}
\label{partial derivative of J_2,2}
\frac{\partial \left[\mathbf{J}_{q}\right]_{2,2}}{\partial {\varrho_{i}}} = \frac{4}{\sigma^{2}} \sum_{l=1}^{L} \sum_{n \neq i}^{N} \mathfrak{Re} \left[je^{j(\textcolor{black}{\varrho_{i}}-\textcolor{black}{\varrho_{n}})} \kappa_{n, i}^{2,2}(l)\right] \;\text{.}
\end{equation}
The Eqs. (\ref{partial derivative of J_1,1})\;-\;(\ref{partial derivative of J_2,2}) are derived in subsection B of Appendix. 

It should be noted that the objective function $f({\boldsymbol{\varrho}})$, \textit{i.e.}, the Cramer-Rao lower bound for the MS position estimate is depend on the unknown parameter $\boldsymbol{\eta}$, and we cannot directly utilize the GDM to optimize the objective function. To address this challenge, we adopt \textit{alternative optimization}, as illustrated in Algorithm 1. More specifically, we start at an initial phase shift vector, and alternately update the estimator of $\boldsymbol{\eta}$ and optimize the phase shift vector $\boldsymbol{\varrho}$, until they converge. The estimation method used in Algorithm 1 is provided in the journal version, and will not be discussed here due to space limitation, and the detailed step of GDM-based reflect beamforming design is illustrated in Algorithm 2.
\begin{algorithm}[!h]
	\begin{spacing}{1.15}
		\caption{The Proposed Alternative Optimization Method}
		\begin{algorithmic}[1]
			\STATE Cramer-Rao lower bound: $f({\boldsymbol{\varrho}})$. 
			\STATE Initialize the phase shift vector ${\boldsymbol{\varrho}}[0]$.
			\WHILE{$\hat{\boldsymbol{\eta}}$ and ${\boldsymbol{\varrho}}$ not converge}
			\STATE $i^{\text{\textit{th}}}$ \textit{iteration}:
			\STATE Estimate the parameter vector $\boldsymbol{\eta}$ and denote as $\hat{\boldsymbol{\eta}}[i]$. 
			\STATE Based on the current iteration estimator $\hat{\boldsymbol{\eta}}[i]$, the phase shift vector is optimized by using the GDM and denote as ${\boldsymbol{\varrho}}[i]$. The GDM used is illustrated in \textcolor{black}{\textit{Algorithm 2}}.
			\ENDWHILE
			\STATE \textbf{Output:} The estimator $\hat{\boldsymbol{\eta}}$ and the reflect beamforming ${\boldsymbol{\varrho}}$.
		\end{algorithmic}
	\end{spacing}
\end{algorithm}

\begin{algorithm}[!h]
	\begin{spacing}{1.15}
		\caption{The Proposed GDM-Based Reflect Beamforming}
		\begin{algorithmic}[1]
			\STATE Objective Function: $f({\boldsymbol{\varrho}})=\mathfrak{Tr}\left(\mathbf{J}_{\boldsymbol{q}}^{-1}\right)$. 
			\STATE \textbf{Input:} The estimator $\hat{\boldsymbol{\eta}}[i]$.
			\STATE Set the number of iterations $j = 0$.
			\STATE Set the stop criterion for the loop: tolerance $\epsilon > 0$.
			\STATE Take the phase shift vector result of the previous iteration as the initialized phase shift vector ${\boldsymbol{\varrho}}[i] = {\boldsymbol{\varrho}[i-1]}$.
			\STATE Compute the objective function value under current ${\boldsymbol{\varrho}}[i]$.
			\WHILE{stop criterion $\Delta f(\boldsymbol{\varrho}[i]) > \epsilon$ is not satisfied}
			\STATE Update the number of iterations $j = j +1$.
			\STATE Choose $-\nabla_{\boldsymbol{\varrho}} f(\boldsymbol{\varrho}[i])$ as the search direction.
			\STATE Choose the step size $t$ via backtracking line search.
			\STATE Update the variable $\boldsymbol{\varrho}[i] := \boldsymbol{\varrho}[i] - t \nabla_{\boldsymbol{\varrho}}f(\boldsymbol{\varrho}[i])$.
			\STATE Compute the objective function difference $\Delta f(\boldsymbol{\varrho}[i])$.
			\ENDWHILE
			\STATE \textbf{Output:} $j$, sub-optimal reflect beamforming $\boldsymbol{\varrho}[i]$.
		\end{algorithmic}
	\end{spacing}
\end{algorithm}

\section{Numerical Results}
This paper focuses on the impact of reflect beamforming on the Cramer-Rao lower bound for the MS position estimate, and aims to minimize it \footnote{
	Some prior works, like \cite{8240645,9124848}, only provide the values of the Cramer-Rao lower bound for the estimate of AoA/AoD. But a small deviation in angle can result in a large deviation in absolute position, and the positioning accuracy can not be reflected directly. Thus, in this section, we present the values of the Cramer-Rao lower bound for the estimate of absolute MS position.
}. The estimation method is provided in the journal version, and will not be discussed here due to space limitation. Thus, we set up the system parameters to be known or have been estimated: the wavelength of mm-wave signal is set up as $0.006$, $\boldsymbol{p} = [0,0,0]^{\mathrm{T}}$, $\boldsymbol{q} = [50,100,0]^{\mathrm{T}}$, and the RIS is a uniform planar array in the vertical plane where $\boldsymbol{s}^{1} = [-20, 50, 20]^{\mathrm{T}}$ and the separation between each reflecting element is $0.1$, all in meters. The propagation gains $h_i \; (i = 1, \cdots, N)$ of $N$ reflection paths are set up as random complex numbers which obey the distribution of $\mathcal{C}\mathcal{N} (0, 1)$.

\subsection{GDM-Based Reflect Beamforming}
We first show the \textit{convergence behaviour} of the proposed GDM-based algorithm in Fig. \ref{fig 2}. In this simulation, we only consider $1$ time slot consumed to transmit the pilot signal. The transmit and receive antenna numbers are both set up as $10$, and the separation equals to half-wavelength $0.003$. The RIS is set up as a $5 \times 5$ uniform planar array, \textit{i.e.}, there exist $25$ reflection paths. The SNR which is defined as $p_{\mathrm{BS}}/(N_{\mathrm{r}}\sigma^2)$ are set up as $30\;\text{dB}$ and $40\;\text{dB}$. (The impact of \textit{time slot number}, \textit{reflecting element number}, and \textit{SNR} on the Cramer-Rao lower bound will be discussed in the next subsection.) The initial phase shift vector is generated randomly with the elements independently drawn from the uniform distribution on $[0,2\pi]$.
\vspace{-2.25 em}
\begin{figure}[htbp]
	\setcounter{figure}{1}
	\centerline{\includegraphics[width = 7.5 cm]{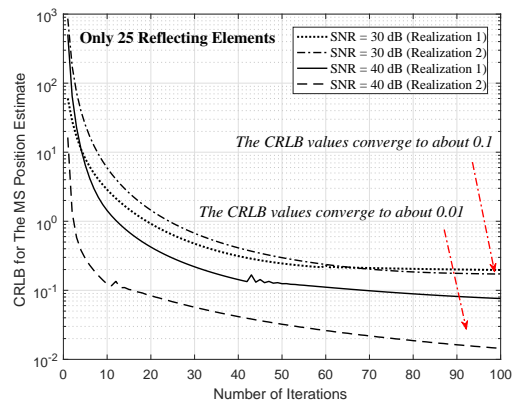}}
	\caption{The convergence behaviour of the proposed GDM-based algorithm.}
	\label{fig 2}
\end{figure}
\vspace{-0.5 em}

The different dotted and solid curves in Fig. \ref{fig 2} are the gradient descent curves with different realizations of propagation gains. It is observed that by searching in the gradient flow direction, the Cramer-Rao lower bound values decrease quickly with the number of iterations. Compared with the case without optimizing the reflect beamforming, there is a significant improvement on Cramer-Rao lower bound when using the proposed algorithm. In addition, the convergence results show that by optimizing the reflect beamforming, the Cramer-Rao lower bound for the estimate of absolute MS position can attain about 0.1 and 0.01, \textit{i.e.}, decimeter-level or even centimeter-level positioning can be achieved by utilizing and optimizing the RIS, which satisfies the requirements of 6G wireless network technology.

\subsection{Impact of Time Slot, Reflecting Element Number, and SNR}
\begin{figure}[htbp]
	\centerline{\includegraphics[width = 7 cm]{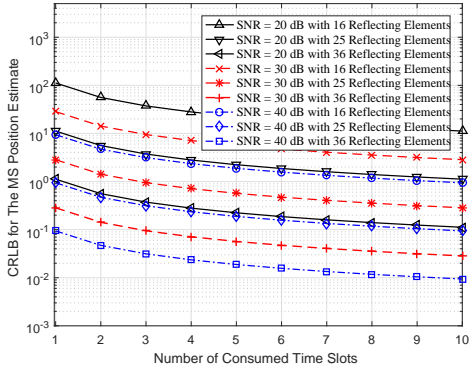}}
	\caption{Cramer-Rao lower bound for the MS position estimate versus time slot with different reflecting element numbers and SNRs.}
	\label{fig 3}
\end{figure}

Next, we show the impact of time slot number, reflecting element number, and SNR in Fig. \ref{fig 3}. In this experiment, we consider the consumed time slot number in the interval of $[1,10]$, the reflecting element number in the set of $\{16, 25, 36\}$, and the SNR in the set of $\{20, 30, 40\}$ [dB]. It is observed that the Cramer-Rao lower bound decreases with the time slot number, the reflecting element number, and the SNR. The decimeter-level or even centimeter-level positioning can be achieved by utilizing the RIS with a large number of reflecting elements. But, it should be noted that the use of a large number of reflecting elements will result in the difficulty of estimating which is also the problem to be solved in the future.
\vspace{-0.75 em}
\vspace{-0.25 em}

\subsection{Impact of MS Position}
\begin{figure}[htbp]
	\centerline{\includegraphics[width = 7 cm]{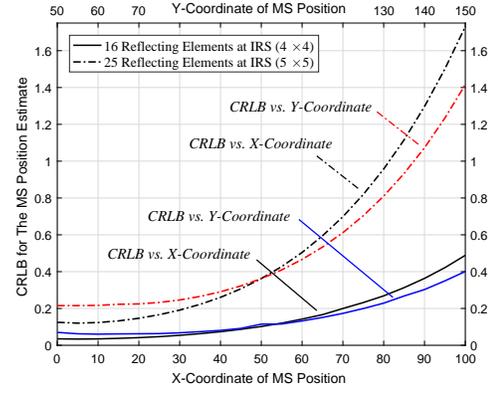}}
	\caption{Cramer-Rao lower bound for the MS position estimate versus $x$ and $y$-coordinates of MS position with different numbers of reflecting elements.}
	\label{fig 4}
\end{figure}

Finally, we show the impact of different MS position on the Cramer-Rao lower bound in Fig. \ref{fig 4}. We consider two cases: one is that the $x$-coordinate of MS position is in the interval of $[0,100]$; Another is that the $y$-coordinate of MS position is in the interval of $[50,150]$. From the figure, we observe that the Cramer-Rao lower bound increases as the MS moves away from the RIS. More accurate localization can be achieved when the MS is closer to the RIS and more reflecting elements are utilized. 

\section{Conclusion}
In this paper, we utilize the advantages of mm-wave signals and large antenna arrays technologies, and introduce the RIS into the system to make the MS positioning more accurate. Our first contribution is to build the three-dimensional RIS-assisted wireless localization system model. After that, we derive the Fisher information matrix and the Cramer-Rao lower bound for the estimate of absolute MS position. Finally, we propose an alternative optimization method and a GDM-based algorithm to make the reflect beamforming design which aims to minimize the Cramer-Rao lower bound. The simulation results show that introducing RIS can make the positioning become possible when the line-of-sight is obstructed, and the proposed algorithm can make the estimation of MS position more accurate. The decimeter-level or even centimeter-level positioning can be achieved by utilizing the RIS with a large number of reflecting elements.

\section*{Appendix}
\subsection{Proof of Eqs. (\ref{submatrix_E}), (\ref{submatrix_F}) and (\ref{submatrix_G})}
By utilizing the geometrical relationship between the azimuth/elevation AoA and the MS position in Eqs. (\ref{geometric relationship between the azimuth AoA and the MS position}) and (\ref{geometric relationship between the elevation AoA and the MS position}), and the derivatives of the inverse functions in Eq. (\ref{derivative of arccos and arctan}),
\begin{equation}
\label{derivative of arccos and arctan}
(\arccos x)^{\prime} = - \frac{1}{\sqrt{1-x^2}} \; \text{,} \; (\arctan x)^{\prime} = \frac{1}{1+x^2} \; \text{,}
\end{equation}
we give the derivatives of $\vartheta_{i}$ and $\varphi_{i}$ with respect to $q_{x}$ as follows:
\begin{equation}
\begin{aligned}
\frac{\partial \varphi_{i}}{\partial q_{x}} & = \frac{\partial \arccos \left[\frac{s_{x}^{i}-q_{x}}{\left\|\boldsymbol{q}_{1:2}-\boldsymbol{s}_{1: 2}^{i}\right\|_{2}}\right]}{\partial q_{x}} \\
& = \frac{\|\boldsymbol{q}_{1:2}-\boldsymbol{s}_{1: 2}^{i}\|_{2} + \left(s_{x}^{i}-q_{x}\right) \frac{1}{2} \frac{1}{\left\|\boldsymbol{q}_{1:2}-\boldsymbol{s}_{1: 2}^{i}\right\|_{2}} 2\left(q_{x}-s^{i}_{x}\right)}{\sqrt{1-\frac{\left(s_{x}^{i}-q_{x}\right)^{2}}{\|\boldsymbol{q}_{1:2}-\boldsymbol{s}_{1: 2}^{i}\|_{2}^{2}}} \; \|\boldsymbol{q}_{1:2}-\boldsymbol{s}_{1: 2}^{i}\|_{2}^{2}} \\
& = \frac{\|\boldsymbol{q}_{1:2}-\boldsymbol{s}_{1: 2}^{i}\|_{2} \left( 1 - \frac{\left(s_{x}^{i}-q_{x}\right)^{2}}{\|\boldsymbol{q}_{1:2}-\boldsymbol{s}_{1: 2}^{i}\|_{2}^{2}} \right)} {\sqrt{1-\frac{\left(s_{x}^{i}-q_{x}\right)^{2}}{\|\boldsymbol{q}_{1:2}-\boldsymbol{s}_{1: 2}^{i}\|_{2}^{2}}} \; \|\boldsymbol{q}_{1:2}-\boldsymbol{s}_{1: 2}^{i}\|_{2}^{2}} \\
& = \frac{\sqrt{1-\frac{\left(s_{x}^{i}-q_{x}\right)^{2}}{\|\boldsymbol{q}_{1:2}-\boldsymbol{s}_{1: 2}^{i}\|_{2}^{2}}}} {\|\boldsymbol{q}_{1:2}-\boldsymbol{s}_{1: 2}^{i}\|_{2}} = \frac{\sin \varphi_{i}} {\|\boldsymbol{q}_{1:2}-\boldsymbol{s}_{1: 2}^{i}\|_{2}} \; \text{,}
\end{aligned}
\end{equation}

\begin{equation}
\begin{aligned}
\frac{\partial \vartheta_{i}}{\partial q_{x}} & = \frac{\partial \arctan \left[\frac{\left\|\boldsymbol{q}_{1:2}-\boldsymbol{s}_{1: 2}^{i}\right\|_{2}}{s_{z}^{i}}\right]}{\partial q_{x}} \\
& = \frac{1}{1+\frac{\left\|\boldsymbol{q}_{1:2}-\boldsymbol{s}_{1: 2}^{i}\right\|_{2}^{2}}{(s_{z}^{i})^{2}}} \frac{\frac{1}{2} \frac{1}{\left\|\boldsymbol{q}_{1:2}-\boldsymbol{s}_{1: 2}^{i}\right\|_{2}} 2\left(q_{x}-s^{i}_{x}\right)s_{z}^{i}}{(s_{z}^{i})^{2}} \qquad \qquad \; \\
& =  \frac{1}{\left\|\boldsymbol{q}_{1:2}-\boldsymbol{s}_{1: 2}^{i}\right\|_{2}^{2}+(s_{z}^{i})^{2}} \frac{\left(q_{x}-s^{i}_{x}\right)s_{z}^{i}}{\left\|\boldsymbol{q}_{1:2}-\boldsymbol{s}_{1: 2}^{i}\right\|_{2}} \\
& = \frac{-s_{z}^{i} \cos \varphi_{i}}{\left\|\boldsymbol{q}_{1:2}-\boldsymbol{s}_{1: 2}^{i}\right\|_{2}^{2}+(s_{z}^{i})^{2}} \; \text{.}
\end{aligned}
\end{equation}

Similarly, we obtain the derivatives of $\vartheta_{i}$ and $\varphi_{i}$ with respect to $q_{y}$ as follows:
\begin{equation}
\frac{\partial \varphi_{i}}{\partial q_{y}} = \frac{\cos \varphi_{i}} {\|\boldsymbol{q}_{1:2}-\boldsymbol{s}_{1: 2}^{i}\|_{2}} \; \text{,} \; \frac{\partial \vartheta_{i}}{\partial q_{y}} = \frac{s_{z}^{i} \sin \varphi_{i}}{\left\|\boldsymbol{q}_{1:2}-\boldsymbol{s}_{1: 2}^{i}\right\|_{2}^{2}+(s_{z}^{i})^{2}}.
\end{equation}

In addition, the absolute MS position $\boldsymbol{q}$ is independent of the propagation gain $h_{i}$, thus we have
\begin{equation}
	\frac{\partial h_{i}}{\partial \boldsymbol{q}_{1:2}} = [0, 0]^{\mathrm{T}}\; \text{.}
\end{equation}

From the above, we complete the derivations of Eqs. (\ref{submatrix_E}), (\ref{submatrix_F}) and (\ref{submatrix_G}) in Section III.

\vspace{0.25 em}
\subsection{Proof of  Eqs. (\ref{partial derivative of J_1,1}), (\ref{partial derivative of J_1,2}), (\ref{partial derivative of J_2,1}) and (\ref{partial derivative of J_2,2})}
Take the partial derivative of the $(1,1)$th entry of the Fisher information matrix $\mathbf{J}_{\boldsymbol{q}}$ with respect to $\textcolor{black}{\varrho_{i}}$ as an example. 

First, we give the partial derivative of $\left[\mathbf{J}_{q}(l)\right]_{1,1}$ as follows:
\begin{equation}
\begin{aligned}
	\frac{\partial \left[\mathbf{J}_{q}(l)\right]_{1,1}}{\partial \textcolor{black}{\varrho_{i}}} & = \frac{\partial \left\lbrace \frac{2}{\sigma^{2}} \sum_{m=1}^{N} \sum_{n=1}^{N} \mathfrak{Re} \left[e^{j(\textcolor{black}{\varrho_{n}}-\textcolor{black}{\varrho_{m}})} \kappa_{m, n}^{1,1}(l)\right] \right\rbrace }{\partial \textcolor{black}{\varrho_{i}}} \\
	& = \frac{\partial \left\lbrace \frac{1}{\sigma^{2}} \sum_{m=1}^{N} \sum_{n=1}^{N} \left[e^{j(\textcolor{black}{\varrho_{n}}-\textcolor{black}{\varrho_{m}})} \kappa_{m, n}^{1,1}(l)\right] \right\rbrace }{\partial \textcolor{black}{\varrho_{i}}} \\
	& +  \frac{\partial \left\lbrace \frac{1}{\sigma^{2}} \sum_{m=1}^{N} \sum_{n=1}^{N} \left[e^{j(\textcolor{black}{\varrho_{m}}-\textcolor{black}{\varrho_{n}})} (\kappa_{m, n}^{1,1}(l))^{*}\right] \right\rbrace }{\partial \textcolor{black}{\varrho_{i}}}\\
	& = { \frac{1}{\sigma^{2}} \sum_{m \neq i}^{N} \left[je^{j(\textcolor{black}{\varrho_{i}}-\textcolor{black}{\varrho_{m}})} \kappa_{m, i}^{1,1}(l)\right] } \\
	& + { \frac{1}{\sigma^{2}} \sum_{n \neq i}^{N} \left[-je^{j(\textcolor{black}{\varrho_{n}}-\textcolor{black}{\varrho_{i}})} \kappa_{i, n}^{1,1}(l)\right] } \\
	& + \frac{1}{\sigma^{2}} \sum_{n \neq i}^{N} \left[je^{j(\textcolor{black}{\varrho_{i}}-\textcolor{black}{\varrho_{n}})} (\kappa_{i, n}^{1,1}(l))^{*}\right] \\
	& +\frac{1}{\sigma^{2}} \sum_{m \neq i}^{N} \left[-je^{j(\textcolor{black}{\varrho_{m}}-\textcolor{black}{\varrho_{i}})} (\kappa_{m, i}^{1,1}(l))^{*}\right] \\
	& = \frac{4}{\sigma^{2}} \sum_{n \neq i}^{N} \mathfrak{Re} \left[je^{j(\textcolor{black}{\varrho_{i}}-\textcolor{black}{\varrho_{n}})} \kappa_{n, i}^{1,1}(l)\right] \;\text{.}
\end{aligned}
\end{equation}
Then, by using Eq. (\ref{eq. 39}), we obtain that
\begin{equation}
	\frac{\partial \left[\mathbf{J}_{q}\right]_{1,1}}{\partial {\varrho_{i}}} = \frac{4}{\sigma^{2}} \sum_{l=1}^{L} \sum_{n \neq i}^{N} \mathfrak{Re} \left[je^{j(\textcolor{black}{\varrho_{i}}-\textcolor{black}{\varrho_{n}})} \kappa_{n, i}^{1,1}(l)\right] \;\text{.}
\end{equation}
The derivations of Eqs. (\ref{partial derivative of J_1,2})\;-\;(\ref{partial derivative of J_2,2}) follow the similar procedures and here we omit them due to space limitation.

From the above, we complete the derivations of Eqs. (\ref{partial derivative of J_1,1})\;-\;(\ref{partial derivative of J_2,2}) in Section IV.

\bibliographystyle{IEEEtran} 
\bibliography{reference}

\begin{thebibliography}{10}
\providecommand{\url}[1]{#1}
\csname url@samestyle\endcsname
\providecommand{\newblock}{\relax}
\providecommand{\bibinfo}[2]{#2}
\providecommand{\BIBentrySTDinterwordspacing}{\spaceskip=0pt\relax}
\providecommand{\BIBentryALTinterwordstretchfactor}{4}
\providecommand{\BIBentryALTinterwordspacing}{\spaceskip=\fontdimen2\font plus
\BIBentryALTinterwordstretchfactor\fontdimen3\font minus
  \fontdimen4\font\relax}
\providecommand{\BIBforeignlanguage}[2]{{%
\expandafter\ifx\csname l@#1\endcsname\relax
\typeout{** WARNING: IEEEtran.bst: No hyphenation pattern has been}%
\typeout{** loaded for the language `#1'. Using the pattern for}%
\typeout{** the default language instead.}%
\else
\language=\csname l@#1\endcsname
\fi
#2}}
\providecommand{\BIBdecl}{\relax}
\BIBdecl

\bibitem{2006.01779}
A.~Bourdoux, A.~N. Barreto, B.~van Liempd, C.~de~Lima, D.~Dardari, D.~Belot,
  E.-S. Lohan, G.~Seco-Granados, H.~Sarieddeen, H.~Wymeersch, J.~Suutala,
  J.~Saloranta, M.~Guillaud, M.~Isomursu, M.~Valkama, M.~R.~K. Aziz,
  R.~Berkvens, T.~Sanguanpuak, T.~Svensson, and Y.~Miao, ``{6G} white paper on
  localization and sensing,'' \emph{arXiv preprint:2006.01779}, June 2020.

\bibitem{6799306}
S.~{Jeong}, O.~{Simeone}, A.~{Haimovich}, and J.~{Kang}, ``Beamforming design
  for joint localization and data transmission in distributed antenna system,''
  \emph{IEEE Transactions on Vehicular Technology}, vol.~64, no.~1, pp. 62--76,
  January 2015.

\bibitem{8240645}
A.~{Shahmansoori}, G.~E. {Garcia}, G.~{Destino}, G.~{Seco-Granados}, and
  H.~{Wymeersch}, ``Position and orientation estimation through millimeter-wave
  {MIMO} in {5G} systems,'' \emph{IEEE Transactions on Wireless
  Communications}, vol.~17, no.~3, pp. 1822--1835, March 2018.

\bibitem{8523631}
Y.~{Wang}, Y.~{Wu}, and Y.~{Shen}, ``Joint spatiotemporal multipath mitigation
  in large-scale array localization,'' \emph{IEEE Transactions on Signal
  Processing}, vol.~67, no.~3, pp. 783--797, February 2019.

\bibitem{8624270}
B.~{Zhou}, A.~{Liu}, and V.~{Lau}, ``Successive localization and beamforming in
  {5G} mm{W}ave {MIMO} communication systems,'' \emph{IEEE Transactions on
  Signal Processing}, vol.~67, no.~6, pp. 1620--1635, March 2019.

\bibitem{Liu2012:Energy}
Y.~Liu, E.~Liu, and R.~Wang, ``Energy efficiency analysis of intelligent
  reflecting surface system with hardware impairments,'' in \emph{2020 IEEE
  Global Communications Conference: Wireless Communications (Globecom2020 WC)},
  Taipei, Taiwan, December 2020.

\bibitem{8811733}
Q.~{Wu} and R.~{Zhang}, ``Intelligent reflecting surface enhanced wireless
  network via joint active and passive beamforming,'' \emph{IEEE Transactions
  on Wireless Communications}, vol.~18, no.~11, pp. 5394--5409, November 2019.

\bibitem{7173440}
Q.~{Nadeem}, A.~{Kammoun}, M.~{Debbah}, and M.~{Alouini}, ``3{D} massive {MIMO}
  systems: Modeling and performance analysis,'' \emph{IEEE Transactions on
  Wireless Communications}, vol.~14, no.~12, pp. 6926--6939, December 2015.

\bibitem{FSSPET}
S.~M. Kay, \emph{Fundamentals of {S}tatistical {S}ignal {P}rocessing:
  Estimation Theory}.\hskip 1em plus 0.5em minus 0.4em\relax Englewood Cliffs,
  NJ: PTR Prentice-Hall, 1993.

\bibitem{SSPDET}
L.~L. Scharf, \emph{Statistical {S}ignal {P}rocess: Detection, Estimation, and
  Time Series Analysis}.\hskip 1em plus 0.5em minus 0.4em\relax Reading, MA:
  Addison-Wesley, 1991.

\bibitem{1519691}
S.~L. {Collier}, ``Fisher information for a complex {G}aussian random variable:
  Beamforming applications for wave propagation in a random medium,''
  \emph{IEEE Transactions on Signal Processing}, vol.~53, no.~11, pp.
  4236--4248, November 2005.

\bibitem{zhang2017matrix}
X.~Zhang, \emph{Matrix analysis and applications}.\hskip 1em plus 0.5em minus
  0.4em\relax Cambridge University Press, 2017.

\bibitem{9124848}
J.~{He}, H.~{Wymeersch}, T.~{Sanguanpuak}, O.~{Silven}, and M.~{Juntti},
  ``Adaptive beamforming design for mmwave {RIS}-aided joint localization and
  communication,'' in \emph{2020 IEEE Wireless Communications and Networking
  Conference Workshops (WCNCW)}, 2020, pp. 1--6.

\end{thebibliography}

\end{document}